\documentclass[twocolumn, prl]{revtex4}
\usepackage{graphicx}
\usepackage{epsfig}
\begin{document}
\title{The Intrinsic Difficulties of Constructing Strongly Correlated States of Lattice Quantum Gases by Connecting Up Pre-engineered  Isolated  Atomic Clusters} 
\author{Tin-Lun Ho}
\affiliation{Department of Physics, The Ohio State University, Columbus, OH 43210}
\date{\today}

\begin{abstract}
Suppose one engineers an artificial antiferromagnet on an array of {\em isolated} wells, and then increases the  tunneling between wells slowly, will the system finally become an {\em equilibrium} antiferromagnet? Here, we show that due to the intrinsic non-adiabaticity at the start of this process, and that the atoms in the initial state  in different wells are completely uncorrelated, the final equilibrium state  will have a temperature $T_{f}$ far above the Neel temperature.  
Constructing other strongly correlated states (with characteristic energy per particle $E^{\ast}$ 
comparable to the hopping or the virtual hopping scale) with the same method 
will suffer the same problem, i.e. $T_{f}>>E^{\ast}$. 
\end{abstract}

\maketitle

At present, there are intense experimental efforts to manufacture strongly correlated states with cold atoms in optical lattices. The success of these attempts will allow one to study many challenging models in condensed matter physics.  These efforts, however,  are exceedingly challenging. The reason is that the energy scales associated with these correlated states are very small, corresponding to temperatures as low as  $10^{-12}$K and beyond\cite{HoZhou}.  Reaching these unprecedentedly low temperatures is a great achievement in itself. 

With the problem of cooling looming at the horizon,  other ideas of achieving strongly correlated states in lattice quantum gases begin to emerge.  One popular idea is to engineer the desired many-body state directly, doing away with cooling completely.  Within this scheme is also the idea of connecting up isolated clusters of atoms, previously engineered into specific quantum states. The hope is that by creating the right kind of clusters, one might achieve the desired {\em equilibrium} many-body states by connecting them up. 
For brevity,  we shall  referred this scheme as {\em ``direct construction method"}. 

For example, to manufacture a resonant valance bond (RVB) liquid or RVB solid, one might start with an array of isolated double wells, each of which contains a singlet pair of spin-1/2 fermions,  as in the experiment of S. Trotzky et.al.\cite{Bloch} 
  One then adjusts the barriers between neighboring double wells  so that at the end one has a cubic lattice with 
one fermion per site.  The hope is that if this process is slow enough, the initial state may evolve adiabatically into a RVB liquid or a RVB solid.   Another example is to manufacture an antiferromagnet using spin-1/2 fermions in an optical lattice. One could first engineer an ``artificial" antiferromagnet consisting of alternating up and down spins in a cubic array of {\em isolated} wells\cite{NIST}. It is tempting to think that  by reducing the barriers between wells sufficiently slowly, the system will turn into an {\em equilibrium} antiferromagnet in a cubic lattice. After all, the initial state is already an ``antiferromagnet" with essentially zero entropy. 

The purpose of this paper is to point out an {\em intrinsic} difficulty of this direct construction scheme. We shall see that the process of connecting up the isolated clusters is intrinsically non-adiabatic. This non-adiabaticity, and the lack of correlations between atoms in different wells in the initial state, will lead to  an energy difference $\sim NE^{\ast}$ 
between the initial and the final equilibrium state, where $N$ is the number of particles, and $E^{\ast}$ is the characteristic energy per particles of the strongly correlated state. 
This excess energy will lead particle fluctuations at the surface, so much so that the temperature $T_{f}$ of the equilibrium state is far above 
 the characteristic energy scale $E^{\ast}$, melting away the correlation one set out to achieve. In the case of anti-ferromagnet, $E^{\ast}$ is the Neel temperature $T_{N}$, and $T_{f}>>T_{N}$.  While our proof does not exclude the possible success of other engineering schemes, it illustrates the kind of problems that must be faced, and the stringent conditions that must be satisfied to reach an {\em equilibrium} strongly correlated state. 
As of now, solving the problem of cooling, (or more correctly, reducing the entropy of the system),    remains essential in achieving strongly correlated states of cold atoms in optical lattices\cite{Squeezing}. 

{\bf I. The initial state and the process of equilibration: } 
Our initial state is an artificial antiferromagnet on a cubic lattice of isolated sites, 
\begin{equation}
|\Psi^{}_{o}\rangle = \prod'_{{\bf r} \epsilon L _{A}}a^{\dagger}_{\bf r \uparrow} \prod'_{{\bf s} \epsilon L _{B}}
a^{\dagger}_{\bf s  \downarrow}|0\rangle, 
\label{Psi} \end{equation}
where $L_{A}$ and $L_{B}$ are the two interpenetrating sub-lattices of a cubic lattice. 
The superscript $'$ means only sites within a radius $R$ from the center are occupied. The density profile therefore has a sharp edge.

The process of connecting up the isolated sites can be described by the following (time dependent) hamiltonian
${\cal H}_{J}= \hat{H}_{J}(U) + \hat{V}+ \hat{\Gamma}$, where $\hat{H}_{J}(U)$ is a 3D Hubbard model describing the internal energy of the a two component Fermi gas in an optical lattice, 
 \begin{equation} 
\hat{H}_{J}(U) = -J\sum_{\bf \langle r, r'\rangle, \sigma }a^{\dagger}_{\bf r \sigma}a^{}_{\bf r' \sigma} +U\sum_{\bf r} n_{\bf r \uparrow}n_{\bf r \downarrow}, 
\label{H} \end{equation}
and $\hat{V}$ is the potential energy of a harmonic trap
\begin{equation}
\hat{V} =  \sum_{\bf r} V_{\bf r}(\omega)  \hat{n}_{\bf r}, \,\,\,\,\,\,\, V_{\bf r}(\omega) = \frac{1}{2} M \omega^2 r^2
\end{equation}
Here, $a^{\dagger}_{\bf r \sigma}$ is the creation operator of a fermion with pseudo-spin $\sigma$ at site ${\bf r}$ of a cubic optical lattice, ($\sigma=\uparrow, \downarrow$),  $\hat{n}_{\bf r \sigma}=a^{\dagger}_{\bf r \sigma}a^{}_{\bf r \sigma}$ is the number of fermions at site ${\bf r}$ with spin $\sigma$, $\hat{n}_{\bf r}= \hat{n}_{\bf r \uparrow}+ \hat{n}_{\bf r \downarrow}$; $U$ is the on-site repulsion,  $J$ is the tunneling integral between nearby sites, and 
$V_{\bf r}= \frac{1}{2} M \omega^2 r^2$ is an harmonic potential with frequency $\omega$. 
The tunneling integral $J$ is time dependent. It grows from 0 to a finite value over a period of time as the lattice height is decreased from infinity to a finite value. $J=0$ corresponds to isolated clusters.   

The term $\hat{\Gamma}$ is the random perturbation due the environment. It can be caused by the  tiny fluctuations in the laser field, or the noise in the current producing the magnetic trap, etc. 
Such perturbations are typically small, and are certainly much weaker than, say potential energy, i.e. 
$\langle \hat{\Gamma}^{2} \rangle^{1/2}<< \langle \hat{V}\rangle$. However,  they can cause de-coherence in quantum evolution. Such de-coherence effect are particularly important at the beginning of the connection process, as the virtual hopping scale $J^2/U$ is vanishingly small and will be dominated by any small perturbation $\hat{\Gamma}$. Because of the strong de-coherence at the beginning, the system at a later time {\em  cannot} be described as simply a quantum mechanical evolution of initial state $|\Psi_{o}\rangle$ under $\hat{H}_{J}^{}(U)$. Rather, one should discuss the properties of the system in terms of density matrix, and how a density matrix associated with the initial state $|\Psi_{o}\rangle$ relaxes to equilibrium. 

Before proceeding, let us quantify the properties of  $|\Psi_{o}\rangle$. The number density of Eq.(\ref{Psi}) can be written as 
\begin{equation}
n^{(o)}_{\bf r} =  \theta (\mu_{o} - V_{\bf r}),  \,\,\,\,\,\,\,\, \mu_{o}  \equiv M\omega^2 R_{o}^2 / 2
\end{equation}
where $\mu_{o}$  is the ``initial" chemical potential, 
 related to total particle number as $N = \sum_{\bf r}n^{(o)}_{\bf r}$, or 
$N= (2\pi/3) [\mu_{o}/ (M\omega^2 a^2/2)]^{3/2}$, where $a$ is the lattice spacing. To eliminate double occupancy, we also have $\mu_{o}<U$.

{\bf II. Formulation of the problem:} 
The evolution of the density matrix will be controlled by : 

\noindent {\bf (A)} $J$ depends exponentially on the barrier height $B$, $J\propto e^{-B}$. As $B$ reduces from infinity to a finite value, $J$ remains exceedingly small for a while. During this period, the tunneling time remains so long that the system can not be in equilibrium and the process will not be adiabatic.\cite{comment}  

\noindent {\bf (B)}  Equilibrium only sets in when $J$ becomes sufficiently large, which only occurs  at a certain time $\tau$ after the infinite barrier was lowered. Because $J$ is changing at a finite rate during this period, energy is being deposited into the system. The determination of this added energy is a complicated problem.

\noindent {\bf (C)} In practice, the time $\tau$ to reach equilibrium must be shorter than the lifetime of the sample, which is limited by  three body collisions and other factors. This imposes a lower bound on the swept rate of $J$, and forces one to face the problem mentioned in {\bf (B)}. 

While both {\bf (A)} and {\bf (B)} are intrinsic, {\bf (C)} comes about because of the specific type of atoms (i.e. the alkalis) used in current experiments. To simplify matters, let us assume that we have atoms with infinite lifetime, and  conduct the following thought experiment: 

\noindent  {\bf (i)} We lower the barrier between wells from infinity to a very large but finite value $J$ over a time interval $t_{o}$.  The tunneling parameter $J$ is sufficiently small so that the tunneling time  $\hbar/J$ is much longer that the switching tim $t_{o}$, i.e. $t_{o}<<\hbar/J$. On the other hand, $J$ is sufficiently large so that even the smallest energy parameter of the system, $J^2/U$, will from now on dominates over the noise term $\hat{\Gamma}$, i.e. $J^2/U >> \gamma$. 
We therefore have following sequence of time scales\cite{comment}, 
\begin{equation}
 \tau_{o} << \hbar/J <<   \hbar/ (J^2/U) << \hbar /\gamma.
 \label{time}\end{equation}

\noindent {\bf (ii)}  After that, we wait for the system to reach equilibrium. Due to the low tunneling rate, this process will take a very long time. However, since the atoms have infinite lifetime, equilibrium will set in eventually.  What we are interested in are the temperature $T$ and the entropy $S$ of the final equilibrium state.  

Since process {\bf (i)} takes place much faster than the tunneling time, the quantum state at the end of this process becomes $|\Psi_{i}\rangle= |\Psi_{o}\rangle + O(\epsilon)$, where $\epsilon = \gamma t_{o}/\hbar <<1$, (see eq.(\ref{time})). 
As a result, the internal energy and potential energy are $E^{int}_{i}= \langle \hat{H}_{J}(U) \rangle_{\Psi_{i}}$  and $V_{i} = \langle \hat{V}\rangle_{\Psi_{i}}$ are given by the value of $|\Psi_{o}\rangle$ to the zeroth order in $\epsilon$, i.e. the total energy at the end of process ${\bf (i)}$ is (or order $O(\epsilon)$), 
\begin{equation}
E_{i} =  E^{int}_{i} + V_{i}= \sum_{\bf r} V_{\bf r}(\omega) n^{(o)}_{\bf r} \,\,\,\,\,\, E^{int}_{i}=0. 
\label{equal}\end{equation}

Next, we note that energy is conserved in process {\bf (ii)}. We then have $E_{i} = E_{f}$, i.e. $E^{int}_{f} + V_{f}^{} = E^{int}_{i} + V_{i}^{}$, or
\begin{equation}
V_{f}^{} - V_{i}^{} = \sum_{\bf r} V_{\bf r}(\omega) (n_{\bf r} - n^{(o)}_{\bf r}) =  E^{int}_{i}  -  E^{int}_{f},  
\label{diff} \end{equation}
where $n_{\bf r}$ and $E^{int}_{f}$  are the density and internal energy of the final equilibrium state, which are functions of the final temperature $T$. Eq.(\ref{diff}) shows that the energy difference between the initial state and the final equilibrium state will lead to particle excitation at the surface, which is the origin of entropy production. Eq.(\ref{diff})  allows us to determine $T$ of the final state once the temperature dependences of $E^{int}$ and $n_{\bf r}$ are known.  Finally, we note that for high barriers and no double occupancy, we have 
 \begin{equation}
 J^2/U << J << \mu_{o}< U. 
 \label{scale} \end{equation}

{\bf III. The energy of the final equilibrium state :}
 We shall assume the final temperature $T$ is less than $U$, for otherwise the system will not be magnetically ordered. For large barrier heights, $U>>J$, the term $J$ in eq.(\ref{H}) can be treated as a perturbation. To the zero order in $J$, we have 
\begin{equation}
n_{\bf r}(T) = \frac{ 2 e^{(\mu - V_{\bf r})/T} +  2 e^{[2(\mu - V_{\bf r}) - U]/T} }{1+ 2 e^{(\mu - V_{\bf r})/T} +  e^{[2(\mu - V_{\bf r}) - U]/T} }, 
\label{n1}\end{equation}
where $\mu$ is chemical potential of the final equilibrium state, determined by $N = \sum_{r} n_{\bf r}$. Since the chemical potential of initial state satisfies  $\mu_{o}<U$, we expect $\mu<U$,  i.e. double occupancy remained disfavored. In this case, (which we shall verify later), we have 
\begin{equation}
n_{\bf r}(T)  \approx  \frac{ 1}{\frac{1}{2}e^{( V_{\bf r} - \mu)/T} + 1}, \,\,\,\,\,\,\,\, N = \sum_{r} n_{\bf r}.
\label{n2}     \end{equation}
Eq.(\ref{n2}) is essentially a step function with size $R$ and a step width $\Delta R$, where 
\begin{equation}
\mu \equiv M\omega^2 R^2/2, \,\,\,\,\,\,\,\, \Delta R = T/(M\omega^2 R). 
\end{equation}
The system can then be divided into following regions: 
\begin{eqnarray}
{\rm Region}\, {\bf 1} :  &r<R-\Delta R/2,   & n_{\bf r}=1  \\
{\rm Region}\, {\bf 2} : &  R- \Delta R/2 < r < \infty,   & 0< n_{\bf r}<1  
\end{eqnarray}
Let $E_{\bf 1}$ and $E_{\bf 2}$ be the total internal energy in region ${\bf 1}$ and ${\bf 2}$, and $\Delta N$ be the number particles in region-{\bf 2}. 
The internal energy $E^{int}_{f}$ of the final state is 
\begin{equation} 
E^{int}_{f}= E_{\bf 1} + E_{\bf 2},  \,\,\,\,
E_{\bf 1} = \epsilon(T) (N-\Delta N), \,\,\,\, E_{\bf 2} = \overline{\epsilon_{2}} \Delta N, 
\label{e1e2} \end{equation}
where $\epsilon(T)$ is the energy per particle of a {\em bulk}  equilibrium anti-ferromagnet at temperature $T$, and $\overline{\epsilon_{2}} $ is the average energy per particle inside the mobile layer. 

To find the energies $ \epsilon$ and $\overline{ \epsilon_{2}}$, we note that in the limit of $U>>J$, the  Hubbard hamiltonian $\hat{H}_{J}^{}(U)($ is reduced to the  $t{\cal J}$ model $\hat{H}_{t{\cal J}}= \hat{\cal T} + \hat{\cal H}_{{\cal J}}$\cite{tJ}, which is a sum of a Heisenberg interaction $\hat{\cal H}_{{\cal J}} =  \frac{1}{2}{\cal J}\sum_{\langle {\bf r,r'}\rangle}{\bf S}_{\bf r}\cdot {\bf S}_{\bf r'}$, ${\cal J}= J^2/U$, and a {\em correlated} hopping term $\hat{\cal T} =  -J\overline{a}^{\dagger}_{\bf r, \sigma}\overline{a}^{}_{\bf r' \sigma}$, where ${\bf S}_{\bf r}$ is the psuedo-spin 1/2 operator of the fermion at site ${\bf r}$, and $\overline{a}^{\dagger}_{\bf r \sigma}$ is a ``correlated" creation operator within the space of no double occupancy.

In region-{\bf 1},  the system has one fermion per site. Hence  $\hat{\cal T}=0$,  and $\hat{H}_{t{\cal J}}$ reduces to the 3D anti-ferromagnetic Heisenberg hamiltonian 
$\hat{\cal H}_{{\cal J}}$.  According to the latest estimate\cite{TN}, the Neel temperature is 
$T_{N}= \gamma {\cal J}$ with $\gamma=0.944$. On dimensional grounds, 
$\epsilon(T) = -  [ \alpha(T/T_{N})] T_{N}$, where $\alpha$ is dimensionless function of $T/T_{N}$. 
Standard spin wave calculation\cite{Kittel} and high temperature series expansion give: 
\begin{equation}
\alpha \rightarrow  \begin{array}{c}  3.58  \\  \gamma^{-2} (3/4)^2T_{N}/T \end{array}    \,\,\, {\rm when } \,\,\, \begin{array}{c} T\rightarrow 0, \\ T>> T_{N}\end{array} . 
\label{high}\end{equation}

In region-{\bf 2}, the mobile layer, the sites are either empty or singly occupied. In this case, $\langle \hat{\cal T}\rangle <0$. Furthermore, due to short range anti-ferromagnetic order, we also have $\langle \hat{\cal H}_{J} \rangle<0$. Hence, we have $\overline{\epsilon_{2}} <0$.  In addition, since $\langle \hat{\cal H}_{J} \rangle\sim (J^2/U) x $, and $\langle \hat{\cal T}\rangle \sim J (1-x) $, where $x$ is the average number of fermion per site in region-{\bf 2}.  Hence we have  $|\overline{\epsilon_{2}}|/J \sim (1-x) + (J/U)x\sim 1$. Combining eq.(\ref{diff}) and (\ref{e1e2}), we have 
\begin{equation}
\sum_{\bf r} V_{\bf r}(\omega) (n_{\bf r} - n^{(o)}_{\bf r}) =   \alpha T_{N} (N-\Delta N)
+ |\overline{\epsilon_{2}}| \Delta N. 
\label{xdiff} \end{equation}
Since $T_N \sim J^2/U<<J$, and $|\overline{\epsilon_{2}}|/J \sim 1$, we have the following
condition on the temperature of the final state, 
\begin{equation}
JN > \sum_{\bf r} V_{\bf r}(\omega) (n_{\bf r} - n^{(o)}_{\bf r}) > \alpha T_{N} (N-\Delta N). 
\label{dddiff} \end{equation}
Note that $n_{\bf r}^{}$ also satisfies
\begin{equation}
N = \sum_{\bf r} n_{\bf r} =   \sum_{\bf r}  n^{(o)}_{\bf r}. 
\label{Ndiff} \end{equation}

{\bf  IV. The temperature of the final state:}
Eq.(\ref{Ndiff}) and eq.(\ref{xdiff}) determine $T$ and $\mu$ of the final state in terms of  the chemical potential $\mu_{o}$ of the initial state,  should the temperature dependence of $\alpha$ and $\overline{\epsilon}_{2}$ are known.      Since eq.(\ref{n2}) reduces to a step function
$\theta(\mu-V_{\bf r})$ as $T\rightarrow 0$, one can apply 
Sommerfeld expansion to evaluate the sums in eq.(\ref{Ndiff}) and eq.(\ref{xdiff}). 
We obtain (see Appendix) 
\begin{equation}
\mu = \mu_{o} \left( 1 - B \frac{T}{\mu_{o}}  - \frac{A}{2}\left( \frac{T}{\mu_{o}}\right)^2 \right),
\label{mu} \end{equation}
\begin{equation}
\sum_{\bf r} V_{\bf r}(\omega) (n_{\bf r} - n^{(o)}_{\bf r})  = N \left( \frac{3A}{2}\right) \left(\frac{T^2}{\mu_{o}} \right)
\label{T} \end{equation}
where $B=0.69$ and $A=1.64$.  
If $\Delta N <<N$, (which we verify later), Eq.(\ref{T}) and (\ref{dddiff}) imply that 
\begin{equation}
J > \left( \frac{3A}{2}\right) \left(\frac{T^2}{\mu_{o}} \right)> \alpha T_{N}. 
\label{integral} \end{equation}
From the temperature dependence of $\alpha$ in eq.(\ref{high}), and the relations in eq.(\ref{scale}), Eq.(\ref{integral}) can be written as
\begin{equation}
1>> \left( \frac{J}{\mu_o} \right)^{1/2} > \frac{T}{\mu_{o}} >  \left( \frac{T_{N}}{\mu_{o}}\right)^{1/2}, \,\,\,\,
\left( \frac{T_{N}}{\mu_{o}}\right)^{1/3}. 
\label{inequality} \end{equation}
where the $(T_{N}/\mu_{o})^{1/2}$ and $(T_{N}/\mu_{o})^{1/3}$ dependence correspond to the low and high temperature behavior of $\alpha$, and we have omitted proportional constants of order 1.  The last inequality of Eq.(\ref{inequality}) shows that 
\begin{equation}
T>\left[  T_{N}\left( \frac{\mu_{o}}{T_{N}}\right)^{1/2} \,\, {\rm or} \,\,\,\, T_{N}\left( \frac{\mu_{o}}{T_{N}}\right)^{2/3}\right]  >> T_{N}
\end{equation}
The last inequality follows from Eq.(\ref{scale}) and the fact that $T_{N} = 0.944(J^2/U)$\cite{Kittel}, which implies $\mu_{o}/T_{N}\approx \mu_{o}/(J^2/U)>>1$.  The final equilibrium state is therefore {\em not} spin ordered.  It is a paramagnetic (rather than anti-ferromagnet) with 
entropy per particle  $\sim {\rm ln}2$ or larger. 
From Eq.(\ref{inequality}), the assumptions $\mu<U$ and $\Delta N<<N$,  from which Eq.(\ref{integral}) is derived, are also easily verified\cite{verification}. 

{\bf V. The condition for ``direct construction" to work:  }
From our analysis, one sees that the severe heating shown above is caused by the energy difference between the initial state and the final equilibrium state, which is of order $NT_{N}$. This excess energy produces particle excitations at the surface,  which generates entropy.   
This energy difference is inherent in the direct construction scheme.  The very fact that one starts with isolated clusters means all the correlations between clusters in the final equilibrium state are all missing in the initial state, which produces an energy difference proportional to the number of clusters,  which is of order $N$.  

One might think that if the traps are tighten immediately after the barriers between neighboring wells are lowered, one will reduce particle fluctuation and hence the final temperature. Mathematically, it means changing Eq.(\ref{diff}) to $\sum_{\bf r}(V_{\bf r}(\omega') n_{\bf r} - V_{\bf r}(\omega)n^{(o)}_{\bf r}) = -E^{int}_{f}$, $ \omega'>\omega$, 
 which will certainly change our conclusions.  This, however, does not work because tightening the trap in the non-adiabatic regime will 
 simply change the initial state to a new one that sees a tighter harmonic trap. In others words, 
Eq.(\ref{diff}) will become $\sum_{\bf r}V_{\bf r}(\omega') (n_{\bf r} - n^{(o)}_{\bf r}) = -E^{int}_{f}$, rather than the expression 
 mentioned above. We are then back to the previous situation, (i.e. Eq.(\ref{diff}) with $\omega$ replaced by $\omega'$), from which the severe heating follows.

While our discussions are for antiferromagnets, our considerations also apply to similar schemes for other strongly correlated states characterized by an energy per particle $E^{\ast}$, which is the energy (or temperature) scale below which the strong correlations between particles emerge.
Since the energy of particle fluctuation at the surface is $\sim (T^2/\mu )N$, in order for final temperature $T$ to be less than $E^{\ast}$,  $T=\zeta E^{\ast}, \zeta <1$, one needs
\begin{equation}
\frac{E^{int}_{i} - E^{int}_{f}}{N} \sim  \frac{\zeta^2 E^{\ast 2}}{\mu}  = E^{\ast}\left(\frac{\zeta^2  E^{\ast}}{\mu}\right)  
\end{equation}
Since $\mu>>E^{\ast}$ in most cases, this condition is very stringent indeed. 

{\bf Appendix: Derivation of  Eq.(\ref{mu}) and (\ref{T})}  
Defining ${\cal E}\equiv V_{\bf r} =M\omega^2 r^2/2$, we have 
$\sum_{\bf r}= \int^{\infty}_{-\infty} {\rm d}{\cal E} D({\cal E})$, where 
$ D({\cal E}) = a^{-3}{\rm d}{\bf r}/d{\cal E} = C\sqrt{{\cal E}}$ for 
${\cal E}>0$  and $ D({\cal E})=0$ for ${\cal E}\leq 0$, 
$C\equiv  \pi  /(M\omega^2 a^2/2)^{3/2}$ and $a$ is the lattice spacing. 
 Eq.(\ref{Ndiff}) and (\ref{xdiff}) can be written as 
\begin{equation}
N=\int^{\infty}_{-\infty} {\rm d}{\cal E} D({\cal E}) n({\cal E})= \int^{\mu_{o}}_{-\infty} {\rm d}{\cal E} D({\cal E}) = (2/3)\mu_{o} D(\mu_{o}) 
\label{AN} \end{equation}
\begin{equation}
V_{f} - V_{i}=\int^{\infty}_{-\infty} {\rm d}{\cal E} D({\cal E}) {\cal E} n({\cal E}) - 
\int^{\mu_{o}}_{-\infty} {\rm d}{\cal E} D({\cal E}) {\cal E}
\label{diff-V} \end{equation}
where $V_{f} - V_{i} = \sum_{\bf r}V_{\bf r}(\omega)(n_{\bf r}- n^{(o)}_{\bf r}) $, and $n({\cal E})=(\frac{1}{2} e^{({\cal E}-\mu)/T}  + 1)^{-1}$.  
Since $n({\cal E})$ is close to but not exactly the Fermi function, we need to generalize the 
usual Sommerfeld expansion. For any function $H({\cal E})$, we have 
\begin{equation}
\int_{-\infty}^{\infty}  {\rm d}{\cal E} H({\cal E}) n({\cal E}) = 
\int_{-\infty}^{{\cal E}^{\ast}}  {\rm d}{\cal E} H({\cal E})  + H'({\cal E}^{\ast}) T^2 A + O(T^3)
\label{Somm}  \end{equation}
where ${\cal E}^{\ast}$ is defined as 
 \begin{equation}
{\cal E}^{\ast} =  \int^{\infty}_{-\infty} {\cal E} \left( - \frac{\partial n}{\partial {\cal E}}\right) {\rm d} {\cal E} = \mu + BT 
\label{East} \end{equation}
with  $B= \int^{\infty}_{-\infty} \frac{{\cal E} - \mu}{T}\left( - \frac{\partial n}{\partial {\cal E}} \right) {\rm d} {\cal E} $$=\int^{\infty}_{-\infty} \frac{ x e^{x}/2 }{\left(  \frac{1}{2}e^{x} + 1\right)^2 }   {\rm d} x = 0.69$, and $A =\int^{\infty}_{-\infty} \frac{ ({\cal E} - {\cal E}^{\ast})^2}{2T^2}\left( - \frac{\partial n}{\partial {\cal E}}\right) {\rm d} {\cal E} $ 
 $= \int^{\infty}_{-\infty} \frac{ (x-B)^2 e^{x}/4}{\left(  \frac{1}{2}e^{x} + 1\right)^2} {\rm d} x = 1.64$.   
 Applying Eq.(\ref{Somm}) to Eq.(\ref{AN}), we have 
 \begin{equation}
({\cal E}^{\ast} - \mu_{o})D({\cal E}^{\ast}) + D'({\cal E}^{\ast})T^2 A =0. 
\label{East-mu} \end{equation}
Together with Eq.(\ref{East}), we have  
 \begin{equation}
{\cal E}^{\ast} =  \mu_{o} - A \frac{D'({\cal E}^{\ast})}{D({\cal E}^{\ast}) } T^2  = \mu + BT. 
\label{EastB} \end{equation}
This gives  eq.(\ref{mu}) upon iteration. 
Likewise, we obtain from Eq.(\ref{diff-V}) and (\ref{Somm})
\begin{equation}
V_{f}-V_{i}=   \int^{{\cal E}^{\ast}}_{\mu_{o}} D({\cal E}){\cal E} 
+ [D({\cal E}^{\ast}){\cal E}^{\ast}]' T^2 A.
 \end{equation}
 Using Eq.(\ref{East-mu}), (\ref{AN}), and (\ref{EastB}), we have 
$V_{f}-V_{i}=  D({\cal E}^{\ast} ) T^2 A  = (3/2) A N(T^2/\mu_{o})  + O(T^3/\mu_{o}^2)$, 
which is Eq.(\ref{T}). 

{\bf Acknowledgement:} 
I would like to thank Bill Phllips and Trey Porto for stimulating discussions that led to this investigation, and Hui Zhai for useful comments on the first draft of the paper. 
This work is supported by NSF Grants DMR0705989, PHY05555576,
 and by DARPA under the Army Research Office Grant Nos. W911NF-07-1-0464, W911NF0710576.


\begin{thebibliography}{99}
\bibitem{HoZhou}  T.L. Ho and Q. Zhou, Phys. Rev. Lett. {\bf 99}, 120404, (2007). 
\bibitem{Bloch}  S. Trotzky et.al., {\em Science}, 319, p.295-299 (2008)
\bibitem{NIST} This can be achieved with the experimental technique in P.J. Lee, et.al. Phys. Rev. Lett. {\bf 99}, 020402 (2007).
\bibitem{Squeezing} A concrete proposal for substantial entropy reduction is given in T.L. Ho and Q. Zhou, arXiv:0808.2652v1. 
\bibitem{comment}  Unlike $J$, which decreases exponentially as barrier height increases, $U$ changes slowly in the high barrier limit. 
\bibitem{tJ} See {\em Interacting Electron and Quantum Magnetism} by Assa Auerbach, 
Springer-Verlag, 1994. 
\bibitem{TN}  J. Oitmaa and Weihong Zheng, Journal of Physics, Condensed Matter {\bf 16} 8653 (2004). 
\bibitem{Kittel} P.61, C. Kittel, Quantum Theory of Solids, second edition, Wiley 1987.  For our caes, we have $z=6, S=1/2$. 
\bibitem{verification}  Eq.(\ref{mu}) shows $\mu< \mu_{o}$. Since $\mu_{o}<U$ (see Eq.(\ref{scale})),  we have $\mu<U$.  Next, we note that  $\Delta N/N = 3\Delta R/R= 3T/M\omega^2 R_{o}^2 = (3/2)(T/\mu)$ $ =  (3/2)(T/\mu_{o}) + O(T/\mu_{o})^{2} <<1$.  Since $T/\mu_{o}<<1$, (eq.(\ref{inequality})), we have $\Delta N/N <<1$. 
\end{thebibliography}
\end{document}